\documentclass[12pt,a4paper]{article}
\usepackage{amsmath}
\usepackage{latexsym}
\usepackage{amssymb}
\usepackage{graphicx,color}
\makeatletter
\def\rddots{\mathinner{\mkern1mu\raise\p@%
    \vbox{\kern7\p@\hbox{.}}\mkern2mu%
    \raise4\p@\hbox{.}\mkern2mu\raise7\p@\hbox{.}\mkern1mu}}
\makeatother
\makeatletter
\def\eqnarray{%
\stepcounter{equation}%
\let\@currentlabel=\theequation
\global\@eqnswtrue
\global\@eqcnt\z@
\tabskip\@centering
\let\\=\@eqncr
$$\halign to \displaywidth\bgroup\@eqnsel\hskip\@centering
$\displaystyle\tabskip\z@{##}$&\global\@eqcnt\@ne
\hfil$\displaystyle{{}##{}}$\hfil
&\global\@eqcnt\tw@$\displaystyle\tabskip\z@{##}$\hfil
\tabskip\@centering&\llap{##}\tabskip\z@\cr}
\makeatother
\begin{document}

\title{\sl Introduction of an Elementary Method to Express 
$\zeta(2n+1)$ in Terms of $\zeta(2k)$ with $k\geq 1$}
\author{
  Kazuyuki FUJII
  \thanks{E-mail address : fujii@yokohama-cu.ac.jp }\quad and \ 
  Tatsuo SUZUKI
  \thanks{E-mail address : suzukita@aoni.waseda.jp ; 
  i027110@sic.shibaura-it.ac.jp }\\
  ${}^{*}$Department of Mathematical Sciences\\
  Yokohama City University\\
  Yokohama, 236--0027\\
  Japan\\
  ${}^{\dagger}$Center for Educational Assistance\\
  Shibaura Institute of Technology\\
  Saitama, 337--8570\\
  Japan\\
  }
\date{}
\maketitle
%
%
%
%
\begin{abstract}
  In this note we give the most elementary method (as far as we know) 
  to express $\zeta(2n+1)$ in terms of $\{\zeta(2k)|k\geq 1\}$. The   
  method is based on only some elementary works by Leonhard Euler, so 
  it is very instructive to non--experts or students.
\end{abstract}

\newpage

%
%
%
%

The zeta function $\zeta(x)$ is defined by
\[
\zeta(x)=\sum_{n=1}^{\infty}\frac{1}{n^{x}} 
\quad \mbox{for}\quad x>1
\]
and we are interested in the (zeta--) values 
$\{\zeta(k)\ |\ k\geq 2\}$
\begin{equation}
\zeta(k)=\sum_{n=1}^{\infty}\frac{1}{n^{k}}.
\end{equation}
The value of $\zeta(2k)$ is well--known by Euler 
\begin{equation}
\zeta(2)=\frac{\pi^{2}}{6},\quad 
\zeta(4)=\frac{\pi^{4}}{90},\quad \mbox{etc.}
\end{equation}
, while that of $\zeta(2k+1)$ is less--known (except for $\zeta(3)$ 
is irrational by Ap\'ery). See \cite{Wol} and its references. 
The textbook \cite{MK} is also recommended. 

\par \noindent
Therefore it is desirable  to express $\zeta(2k+1)$ in terms of 
$\{\zeta(2k)|k\geq 1\}$ like
\begin{equation}
\label{eq:expression}
\zeta(2k+1)=c_{0}+\sum_{n=1}^{\infty}c_{n}\zeta(2n)
\end{equation}
where $\{c_{n}|n\geq 0\}$ are constants.

In this note we revisit the problem and give a method by use of 
only a few elementary works by Euler. 
It is most elementary \footnote{we don't know the precise 
definition of ``most elementary"} as far as we know.

Let us start by listing well--known works by Euler :
\begin{eqnarray*}
\label{Euler}
&&(a)\ \ \sin x=\frac{e^{ix}-e^{-ix}}{2i} \\
&&(b)\ \ \sin x=x\prod_{n=1}^{\infty}
                \left(1-\frac{x^{2}}{n^{2}\pi^{2}}\right) \\
&&(c)\ \ \int_{0}^{\frac{\pi}{2}}x\log(\sin x)dx=
         -\frac{\pi^{2}}{8}\log 2+\frac{7}{16}\zeta(3)
\end{eqnarray*}
Though the equation (c) may be not popular among the Euler's works 
(which are huge !\ \cite{WD}) it is important and interesting 
enough as shown in the following.

For the sake of non--experts who are interested in the zeta values 
we show our method with simple example (c), which is very instructive. 
Namely, we calculate the integral
\begin{equation}
\label{eq:first-integral}
\int_{0}^{\frac{\pi}{2}}x\log(\sin x)dx
\end{equation}
in two ways by use of (a) and (b). 

Here we list some well--known integrals related to 
(\ref{eq:first-integral})
\begin{eqnarray*}
\int_{0}^{\frac{\pi}{2}}\log(\sin x)dx&=&-\frac{\pi}{2}\log 2 \\
\int_{0}^{\frac{\pi}{2}}\sin x\log(\sin x)dx&=&\log 2-1 \\
\int_{0}^{\frac{\pi}{2}}x\log{x}dx&=&
\frac{\pi^{2}}{8}\log\left(\frac{\pi}{2}\right)-\frac{\pi^{2}}{16}
\end{eqnarray*}
for the convenience of readers.

\vspace{5mm}
\par \noindent
(I)\ \ \underline{Calculation by use of (a)}

By (a)
\begin{eqnarray}
\label{eq:log-expansion}
\log(\sin x)
&=&\log(\frac{e^{ix}-e^{-ix}}{2i})
=\log(e^{ix}-e^{-ix})-\log(2i)
=\log\{e^{ix}(1-e^{-2ix})\}-\log(2i)  \nonumber \\
&=&ix-\log(2i)+\log(1-e^{-2ix})  
=ix-\log(2i)-\sum_{n=1}^{\infty}\frac{e^{-2inx}}{n},
\end{eqnarray} 
where we have used the Taylor expansion
\[
\log(1-z)=-\sum_{n=1}^{\infty}\frac{z^{n}}{n}
\quad\mbox{for}\quad |z|<1.
\]
Then
\begin{eqnarray*}
\int_{0}^{\frac{\pi}{2}}x\log(\sin x)dx
&=&
i\int_{0}^{\frac{\pi}{2}}x^{2}dx
-\log(2i)\int_{0}^{\frac{\pi}{2}}xdx
-\sum_{n=0}^{\infty}\frac{1}{n}\int_{0}^{\frac{\pi}{2}}xe^{-2inx}dx \\
&=&
i\frac{\pi^{3}}{24}-\frac{\pi^{2}}{8}\log(2i)
-\sum_{n=1}^{\infty}\frac{1}{n}\int_{0}^{\frac{\pi}{2}}xe^{-2inx}dx
\end{eqnarray*} 
and
\begin{eqnarray*}
\int_{0}^{\frac{\pi}{2}}xe^{-2inx}dx
&=&
\left[\frac{e^{-2inx}}{-2in}x\right]_{0}^{\frac{\pi}{2}}+
\frac{1}{2in}\int_{0}^{\frac{\pi}{2}}e^{-2inx}dx \\
&=&
\frac{i\pi}{4n}e^{-i\pi n}-\frac{1}{4n^{2}}(1-e^{-i\pi n})
=
\frac{i\pi}{4n}(-1)^{n}-\frac{1}{4n^{2}}(1-(-1)^{n}).
\end{eqnarray*}
Therefore
\begin{eqnarray*}
\int_{0}^{\frac{\pi}{2}}x\log(\sin x)dx
&=&
i\frac{\pi^{3}}{24}-\frac{\pi^{2}}{8}\log(2i)
-\sum_{n=1}^{\infty}\frac{1}{n}
\left\{\frac{i\pi}{4n}(-1)^{n}-\frac{1}{4n^{2}}(1-(-1)^{n})\right\} \\
&=&
i\frac{\pi^{3}}{24}-\frac{\pi^{2}}{8}\log(2i)
-\frac{i\pi}{4} \sum_{n=1}^{\infty}\frac{(-1)^{n}}{n^{2}}  
+\frac{1}{4}\sum_{n=1}^{\infty}\frac{1-(-1)^{n}}{n^{3}}   \\
&=&
i\frac{\pi^{3}}{24}-\frac{\pi^{2}}{8}\log(2i)
+\frac{i\pi}{4} \sum_{n=1}^{\infty}\frac{(-1)^{n-1}}{n^{2}}  
+\frac{1}{2}\sum_{n=1}^{\infty}\frac{1}{(2n-1)^{3}}. 
\end{eqnarray*} 
Since it is easy to see
\begin{eqnarray*}
\sum_{n=1}^{\infty}\frac{(-1)^{n-1}}{n^{2}}
=
\sum_{n=1}^{\infty}\frac{1}{(2n-1)^{2}}-
\sum_{n=1}^{\infty}\frac{1}{(2n)^{2}}
=
\sum_{n=1}^{\infty}\frac{1}{n^{2}}-
2\sum_{n=1}^{\infty}\frac{1}{(2n)^{2}}
=
\frac{1}{2}\zeta(2)
\end{eqnarray*}
and
\begin{eqnarray*}
\sum_{n=1}^{\infty}\frac{1}{(2n-1)^{3}}
=
\sum_{n=1}^{\infty}\frac{1}{n^{3}}-
\sum_{n=1}^{\infty}\frac{1}{(2n)^{3}}
=
\frac{7}{8}\zeta(3)
\end{eqnarray*}
we obtain
\begin{eqnarray}
\label{eq:result-1}
\int_{0}^{\frac{\pi}{2}}x\log(\sin x)dx
&=&
i\frac{\pi^{3}}{24}-\frac{\pi^{2}}{8}\log(2i)
+\frac{i\pi}{4}\times \frac{1}{2}\zeta(2) 
+\frac{1}{2}\times \frac{7}{8}\zeta(3) \nonumber \\
&=&
i\frac{\pi^{3}}{24}-
\frac{\pi^{2}}{8}(\log 2+\log(i))
+\frac{i\pi}{8}\zeta(2)+\frac{7}{16}\zeta(3) \nonumber \\
&=&
-\frac{\pi^{2}}{8}\log 2+\frac{7}{16}\zeta(3)+
i\left(
\frac{\pi^{3}}{24}-\frac{\pi^{3}}{16}+\frac{\pi}{8}\zeta(2)
\right)  \nonumber \\
&=&
-\frac{\pi^{2}}{8}\log 2+\frac{7}{16}\zeta(3)+
i\frac{\pi}{8}\left(\zeta(2)-\frac{\pi^{2}}{6}\right),
\end{eqnarray} 
where we have used the principal value
\[
\log(i)=\mbox{Log}(e^{\frac{i\pi}{2}})=i\frac{\pi}{2}.
\]
As a result we have $\zeta(2)=\frac{\pi^{2}}{6}$ {\bf automatically} 
and the equation (c).

\vspace{5mm}
A comment is in order.\ \ Our proof is not rigorous in the 
mathematical sense because the convergence radius of 
(\ref{eq:log-expansion}) is ignored. On the other hand, it is very 
clear why $\zeta(3)$ appears in the process of calculation.

\vspace{5mm}\noindent
(II)\ \ \underline{Calculation by use of (b)}

By (b)
\begin{equation}
\log(\sin x)=\log x+\sum_{n=1}^{\infty}
\log\left(1-\frac{x^{2}}{n^{2}\pi^{2}}\right),
\end{equation}
so
\begin{eqnarray*}
\int_{0}^{\frac{\pi}{2}}x\log(\sin x)dx
&=&
\int_{0}^{\frac{\pi}{2}}x\log xdx+
\sum_{n=1}^{\infty}\int_{0}^{\frac{\pi}{2}}
x\log\left(1-\frac{x^{2}}{n^{2}\pi^{2}}\right)dx \\
&=&
\frac{\pi^{2}}{8}\log\left(\frac{\pi}{2}\right)-\frac{\pi^{2}}{16}+
\sum_{n=1}^{\infty}\int_{0}^{\frac{\pi}{2}}
x\log\left(1-\frac{x^{2}}{n^{2}\pi^{2}}\right)dx.
\end{eqnarray*} 
Let us calculate the last term. 
By the change of variables ($x\ \longrightarrow\ n\pi\sqrt{x}$)
\[
\int_{0}^{\frac{\pi}{2}}
x\log\left(1-\frac{x^{2}}{n^{2}\pi^{2}}\right)dx
=
\frac{n^{2}\pi^{2}}{2}\int_{0}^{\frac{1}{4n^{2}}}\log(1-x)dx
\] 
and using the Taylor expansion
\[
\log(1-x)=-\sum_{n=1}^{\infty}\frac{x^{n}}{n}
\]
we obtain
\begin{eqnarray*}
\int_{0}^{\frac{\pi}{2}}
x\log\left(1-\frac{x^{2}}{n^{2}\pi^{2}}\right)dx
&=&
\frac{n^{2}\pi^{2}}{2}
\left\{
-\sum_{k=1}^{\infty}\frac{1}{k}
\int_{0}^{\frac{1}{4n^{2}}}x^{k}dx
\right\} \\
&=&
-\frac{n^{2}\pi^{2}}{2}
\sum_{k=1}^{\infty}\frac{1}{k(k+1)}\frac{1}{4^{k+1}n^{2k+2}}
=
-\frac{\pi^{2}}{8}
\sum_{k=1}^{\infty}\frac{1}{k(k+1)2^{2k}}\frac{1}{n^{2k}}.
\end{eqnarray*} 
As a result we have
\begin{eqnarray}
\int_{0}^{\frac{\pi}{2}}x\log(\sin x)dx
&=&
\frac{\pi^{2}}{8}\log\left(\frac{\pi}{2}\right)-\frac{\pi^{2}}{16}
-\frac{\pi^{2}}{8}
\sum_{k=1}^{\infty}\frac{1}{k(k+1)2^{2k}}
\left(\sum_{n=1}^{\infty}\frac{1}{n^{2k}}\right)  \nonumber \\
&=&
\frac{\pi^{2}}{8}\log\left(\frac{\pi}{2}\right)-\frac{\pi^{2}}{16}
-\frac{\pi^{2}}{8}\sum_{k=1}^{\infty}\frac{\zeta(2k)}{k(k+1)2^{2k}}.
\end{eqnarray} 

\vspace{5mm}
By comparing (I) with (II) we obtain the expression of 
$\zeta(3)$
\begin{equation}
\label{eq:Our-expression}
\zeta(3)=\frac{2\pi^{2}}{7}
\left\{
\log\pi-\frac{1}{2}-\sum_{n=1}^{\infty}\frac{\zeta(2n)}{n(n+1)2^{2n}}
\right\}.
\end{equation}
However, our expression (\ref{eq:Our-expression}) is of course not new, 
see for example \cite{Sugi} or \cite{Sato}.

By the way, the expression by Euler is different from ours :
\begin{equation}
\label{eq:Euler-expression}
\zeta(3)=\frac{\pi^{2}}{7}
\left\{1-4\sum_{n=1}^{\infty}\frac{\zeta(2n)}{(2n+1)(2n+2)2^{2n}}\right\}
,
\end{equation}
see \cite{Wol}. Therefore these two expressions give the (interesting) 
equation 
\begin{equation}
\log{\pi}=1+\sum_{n=1}^{\infty}\frac{\zeta(2n)}{n(2n+1)2^{2n}}
\ \Longleftrightarrow\ 
\log\left(\frac{\pi}{e}\right)=
\sum_{n=1}^{\infty}\frac{\zeta(2n)}{n(2n+1)2^{2n}}.
\end{equation}

\vspace{5mm}
In the following we generalize the method above to obtain 
the equation (\ref{eq:expression}). 
For that purpose we consider the integral
\begin{equation}
\label{eq:general-integral}
\int_{0}^{\frac{\pi}{2}}x^{2l-1}\log(\sin x)dx
\quad\mbox{for}\quad l\geq 1.
\end{equation}
It may be reasonable to call this  the {\bf Euler integral}. 

We calculate (\ref{eq:general-integral}) in two ways by use of 
(a) and (b).

\vspace{5mm}
\par \noindent
(I$'$)\ \ \underline{Calculation by use of (a)}

In a similar way in (I) it is easy to see
\begin{eqnarray*}
\int_0^{\frac{\pi}{2}} x^{2l-1} \log(\sin x)dx
&=&
i \int_0^{\frac{\pi}{2}}x^{2l}dx-\log(2i)\int_0^{\frac{\pi}{2}}x^{2l-1}dx
-\sum_{n=1}^{\infty}\frac{1}{n}\int_0^{\frac{\pi}{2}}
x^{2l-1}e^{-2inx}dx  \\
&=&
i\frac{(\frac{\pi}{2})^{2l+1}}{2l+1}
-\left(\log 2+ i\frac{\pi}{2}\right)\frac{(\frac{\pi}{2})^{2l}}{2l}
-\sum_{n=1}^{\infty}\frac{1}{n}\int_0^{\frac{\pi}{2}}
x^{2l-1} e^{-2inx}dx  \\
&=&
i\frac{(\frac{\pi}{2})^{2l+1}}{2l+1}-i\frac{(\frac{\pi}{2})^{2l+1}}{2l}
-\frac{(\frac{\pi}{2})^{2l}}{2}\log 2
-\sum_{n=1}^{\infty}\frac{1}{n}\int_0^{\frac{\pi}{2}}
x^{2l-1} e^{-2inx}dx  \\
&=&
-i\frac{(\frac{\pi}{2})^{2l+1}}{2l(2l+1)}-
\frac{(\frac{\pi}{2})^{2l}}{2l}\log 2
-\sum_{n=1}^{\infty}\frac{1}{n}\int_0^{\frac{\pi}{2}}
x^{2l-1} e^{-2inx}dx.
\end{eqnarray*}
In order to calculate the last term (which is not so easy) we make use 
of the trick. From
\[
\int_0^{\frac{\pi}{2}}e^{-2inx}dx
=\frac{e^{-i\pi n}-1}{-2i n}
=-\frac{1}{2i}\{(e^{-i\pi n}-1)n^{-1}\}
\]
we differentiate the equation above $2l-1$ times with respect to $n$ 
\begin{eqnarray*}
&&(-2i)^{2l-1}
\int_0^{\frac{\pi}{2}}x^{2l-1}e^{-2inx}dx
=-\frac{1}{2i}
\left( \frac{d}{dn} \right)^{2l-1}
\{ (e^{-i\pi n}-1)n^{-1} \} \\
&=&
-\frac{1}{2i}
\left\{ \sum_{j=0}^{2l-2} \frac{(2l-1)!}{j!(2l-1-j)!}(-i\pi)^{2l-1-j}
 e^{-i\pi n}\cdot (-1)^j j! n^{-j-1}
+ (e^{-i\pi n}-1)(-1)^{2l-1}(2l-1)!n^{-2l} \right\} \\
&=&
\frac{(2l-1)!}{-2i}
\left\{ \sum_{j=0}^{2l-2} \frac{(i\pi)^{2l-1-j}}{(2l-1-j)!}
\frac{(-1)^{n-1}}{n^{j+1}}
+ \frac{1-(-1)^n}{n^{2l}} \right\}
\end{eqnarray*}
where we have used the Leibniz's rule of differentiation, so
\begin{eqnarray*}
\int_0^{\frac{\pi}{2}}x^{2l-1}e^{-2inx}dx
&=&
\frac{(2l-1)!}{(-2i)^{2l}}
\left\{ \sum_{j=0}^{2l-2} \frac{(i\pi)^{2l-1-j}}{(2l-1-j)!}
\frac{(-1)^{n-1}}{n^{j+1}}
+ \frac{1-(-1)^n}{n^{2l}} \right\} \\
&=&
\frac{(-1)^{l}(2l-1)!}{2^{2l}}
\left\{ \sum_{j=0}^{2l-2} \frac{(i\pi)^{2l-1-j}}{(2l-1-j)!}
\frac{(-1)^{n-1}}{n^{j+1}}
+ \frac{1-(-1)^n}{n^{2l}} \right\}.
\end{eqnarray*}

By noting
\[
\sum_{n=1}^{\infty}\frac{(-1)^{n-1}}{n^{k}}=
\left(1-\frac{1}{2^{k-1}}\right)\zeta(k),
\quad
\sum_{n=1}^{\infty}\frac{1-(-1)^{n}}{n^{k}}=
2\left(1-\frac{1}{2^{k}}\right)\zeta(k)
\]
we obtain

\begin{eqnarray*}
&&
\int_0^{\frac{\pi}{2}} x^{2l-1} \log(\sin x)dx  \\
&=&
-i\frac{(\frac{\pi}{2})^{2l+1}}{2l(2l+1)}
-\frac{(\frac{\pi}{2})^{2l}}{2l}\log 2
+\frac{(-1)^{l-1}(2l-1)!}{2^{2l}}
\left\{ \sum_{j=0}^{2l-2} \frac{(i\pi)^{2l-1-j}}{(2l-1-j)!}
\sum_{n=1}^{\infty}\frac{(-1)^{n-1}}{n^{j+2}}
+\sum_{n=1}^{\infty} \frac{1-(-1)^n}{n^{2l+1}} \right\} \\
&=&
-i\frac{(\frac{\pi}{2})^{2l+1}}{2l(2l+1)}
-\frac{(\frac{\pi}{2})^{2l}}{2l}\log 2 \\
&& \hspace{1cm}
+\frac{(-1)^{l-1}(2l-1)!}{2^{2l}}
\left\{ \sum_{j=0}^{2l-2} \frac{(i\pi)^{2l-1-j}}{(2l-1-j)!}
\left( 1-\frac{1}{2^{j+1}} \right) \zeta(j+2)
+2\left( 1-\frac{1}{2^{2l+1}} \right) \zeta(2l+1) \right\} \\
&& \\
&&
(\mbox{by dividing the summation into}\ j=2k \ (k=0, \cdots, l-1)\ 
\mbox{and} \ j=2k-1 \ (k=1, \cdots, l-1)\ ) \\
&& \\
&=&
-i\frac{(\frac{\pi}{2})^{2l+1}}{2l(2l+1)}
-\frac{(\frac{\pi}{2})^{2l}}{2l}\log 2 \\
&& \hspace{1cm} 
+\frac{(-1)^{l-1}(2l-1)!}{2^{2l}}
\sum_{k=0}^{l-1} \frac{(i\pi)^{2l-1-2k}}{(2l-1-2k)!}
\left( 1-\frac{1}{2^{2k+1}} \right) \zeta(2k+2) \\
&& \hspace{1cm} 
+\frac{(-1)^{l-1}(2l-1)!}{2^{2l}}
\left\{ \sum_{k=1}^{l-1} \frac{(i\pi)^{2(l-k)}}{(2l-2k)!}
\left( 1-\frac{1}{2^{2k}} \right) \zeta(2k+1)
+\frac{2^{2l+1}-1}{2^{2l}}\zeta(2l+1) \right\} \\
&=&
-\frac{(\frac{\pi}{2})^{2l}}{2l}\log 2
+\frac{(-1)^{l-1}(2l-1)!}{2^{2l}}
\sum_{k=1}^{l-1} \frac{(-1)^{l-k}\pi^{2(l-k)}}{(2(l-k))!}
\left( 1-\frac{1}{2^{2k}} \right) \zeta(2k+1) \\
&& \hspace{2cm}\ \  
+(-1)^{l-1}(2l-1)!
\frac{2^{2l+1}-1}{2^{4l}}\zeta(2l+1) \\
&&
+i \left\{ 
-\frac{(\frac{\pi}{2})^{2l+1}}{2l(2l+1)}
+\frac{(-1)^{l-1}(2l-1)!}{2^{2l}}
\sum_{k=0}^{l-1} \frac{(-1)^{l-k-1}\pi^{2(l-k)-1}}{(2(l-k)-1)!}
\left( 1-\frac{1}{2^{2k+1}} \right) \zeta(2k+2) \right\} \\
&=&
-\frac{(\frac{\pi}{2})^{2l}}{2l}\log 2
+\frac{(2l-1)!}{2^{2l}}
\sum_{k=1}^{l-1} \frac{(-1)^{k-1}\pi^{2(l-k)}}{(2(l-k))!}
\left( 1-\frac{1}{2^{2k}} \right) \zeta(2k+1) \\
&& \hspace{2cm}\ \  
+(-1)^{l-1}(2l-1)!
\frac{2^{2l+1}-1}{2^{4l}}\zeta(2l+1) \\
&&
+i \left\{ 
-\frac{(\frac{\pi}{2})^{2l+1}}{2l(2l+1)}
+\frac{(2l-1)!}{2^{2l}}
\sum_{k=0}^{l-1} \frac{(-1)^{k}\pi^{2(l-k)-1}}{(2(l-k)-1)!}
\left( 1-\frac{1}{2^{2k+1}} \right) \zeta(2k+2) \right\}.
\end{eqnarray*}

From this equation the imaginary part must be zero, so we have
\[
\frac{(2l-1)!}{2^{2l}}
\sum_{k=0}^{l-1} \frac{(-1)^{k}\pi^{2(l-k)-1}}{(2(l-k)-1)!}
\left( 1-\frac{1}{2^{2k+1}} \right) \zeta(2k+2)
-\frac{(\frac{\pi}{2})^{2l+1}}{2l(2l+1)}
=0
\quad (l=1,2,\cdots).
\]
Here, if we asuume
\begin{equation}
\zeta(0)=-\frac{1}{2}
\end{equation}
then the equation above is rewritten in a compact form
\[
\sum_{k=-1}^{l-1} \frac{(-1)^{k}\pi^{2(l-k)-1}}{(2(l-k)-1)!}
\left( 1-\frac{1}{2^{2k+1}} \right) \zeta(2k+2)=0
\quad (l=1,2,\cdots)
\]
or ($k\ \longrightarrow\ k-1$)
\[
\sum_{k=0}^{l} \frac{(-1)^{k-1}\pi^{2(l-k)+1}}{(2(l-k)+1)!}
\left( 1-\frac{1}{2^{2k-1}} \right) \zeta(2k)=0
\quad (l=1,2,\cdots).
\]
From this we have the recurrent relation

\par \noindent
{\bf Result 1}\quad For $l=1,2,\cdots$
\begin{equation}
\label{eq:even}
\zeta(2l)=
\frac{2^{2l-1}}{2^{2l-1}-1}
\sum_{k=0}^{l-1} \frac{(-1)^{l+k-1}\pi^{2(l-k)}}{(2(l-k)+1)!}
\left( 1-\frac{1}{2^{2k-1}} \right) \zeta(2k).
\end{equation}
Let us list some examples : 
\[
\zeta(2)=\frac{\pi^{2}}{6}, \quad
\zeta(4)=\frac{\pi^{4}}{90}, \quad \mbox{etc.}
\]

Next, from the real part of the equation we have

\par \noindent
{\bf Result 2}\quad For $l=1,2,\cdots$
\begin{eqnarray}
\label{eq:2}
\int_0^{\frac{\pi}{2}} x^{2l-1} \log(\sin x)dx
&=&
-\frac{(\frac{\pi}{2})^{2l}}{2l}\log 2 
+\frac{(2l-1)!}{2^{2l}}
\sum_{k=1}^{l-1} \frac{(-1)^{k-1}\pi^{2(l-k)}}{(2(l-k))!}
\left( 1-\frac{1}{2^{2k}} \right) \zeta(2k+1) \nonumber \\
&& \hspace{2cm}\ \  
+(-1)^{l-1}(2l-1)!
\frac{2^{2l+1}-1}{2^{4l}}\zeta(2l+1). \label{eqn:1}
\end{eqnarray}
Let us list some examples : 
\begin{eqnarray*}
\int_0^{\frac{\pi}{2}}x\log(\sin x)dx
&=&
-\frac{{\pi}^2}{8}\log 2+\frac{7}{16}\zeta(3), \\
\int_0^{\frac{\pi}{2}}x^3\log(\sin x)dx
&=&
-\frac{{\pi}^4}{64}\log 2+\frac{9{\pi}^2}{64}\zeta(3)
-\frac{93}{128}\zeta(5),\quad \mbox{etc.}
\end{eqnarray*}

\vspace{5mm}
\par \noindent
(II$'$)\ \ \underline{Calculation by use of (b)}

In a similar way in (II) it is easy to see
\begin{eqnarray*}
\int_0^{\frac{\pi}{2}}x^{2l-1} \log(\sin x)dx
&=&\int_0^{\frac{\pi}{2}} x^{2l-1} \log x dx
+\sum_{n=1}^{\infty}
\int_0^{\frac{\pi}{2}}x^{2l-1} 
\log \left( 1-\frac{x^2}{n^2 \pi^2} \right) dx \nonumber \\
&=&
\frac{(\frac{\pi}{2})^{2l}}{2l}\log \left( \frac{\pi}{2} \right)
-\frac{(\frac{\pi}{2})^{2l}}{(2l)^2}
+\sum_{n=1}^{\infty}
\frac{(n^2 \pi^2)^l}{2}
\int_0^{\frac{1}{4n^2}} t^{l-1} \log(1-t) dt \nonumber \\
&=&
\frac{(\frac{\pi}{2})^{2l}}{2l}\log \left( \frac{\pi}{2} \right)
-\frac{(\frac{\pi}{2})^{2l}}{(2l)^2}
-\sum_{n=1}^{\infty}
\frac{(n^2 \pi^2)^l}{2}
\sum_{k=1}^{\infty}
\frac{1}{k}\int_0^{\frac{1}{4n^2}} t^{k+l-1} dt \nonumber \\
&=&
\frac{(\frac{\pi}{2})^{2l}}{2l}\log \left( \frac{\pi}{2} \right)
-\frac{(\frac{\pi}{2})^{2l}}{(2l)^2}
-\sum_{n=1}^{\infty}
\frac{(n^2 \pi^2)^l}{2}
\sum_{k=1}^{\infty}\frac{1}{k(k+l)(4n^2)^{k+l}} \nonumber \\
&=&
\frac{(\frac{\pi}{2})^{2l}}{2l}\log \left( \frac{\pi}{2} \right)
-\frac{(\frac{\pi}{2})^{2l}}{(2l)^2}
-\sum_{n=1}^{\infty}
\frac{n^{2l} \pi^{2l}}{2}
\sum_{k=1}^{\infty}\frac{1}{k(k+l)2^{2(k+l)}n^{2k}n^{2l}} \nonumber \\
&=&
\frac{(\frac{\pi}{2})^{2l}}{2l}\log \left( \frac{\pi}{2} \right)
-\frac{(\frac{\pi}{2})^{2l}}{(2l)^2}
-\left( \frac{\pi}{2} \right)^{2l}\cdot 
\frac{1}{2}
\sum_{k=1}^{\infty}\frac{\zeta(2k)}{k(k+l)2^{2k}} \nonumber \\
&=&
\left( \frac{\pi}{2} \right)^{2l}
\left\{ 
\frac{1}{2l}(\log \pi - \log 2)
-\frac{1}{(2l)^2}
-\frac{1}{2}
\sum_{k=1}^{\infty}\frac{\zeta(2k)}{k(k+l)2^{2k}} 
\right\}.
\end{eqnarray*}
Therefore we have

\par \noindent
{\bf Result 3}\quad For $l=1,2,\cdots$
\begin{equation}
\label{eq:3}
\int_0^{\frac{\pi}{2}}x^{2l-1} \log(\sin x)dx
=
\left( \frac{\pi}{2} \right)^{2l}
\left\{ 
\frac{1}{2l}(\log \pi - \log 2)
-\frac{1}{(2l)^2}
-\frac{1}{2}
\sum_{k=1}^{\infty}\frac{\zeta(2k)}{k(k+l)2^{2k}} 
\right\}.
\end{equation}

\vspace{5mm}
By comparing (\ref{eq:2}) with (\ref{eq:3}) we have the main result

\par \noindent
{\bf Result 4 (Main)}\quad For $l=1,2,\cdots$
\begin{eqnarray}
\label{eq:main}
\zeta(2l+1)
&=&
\frac{(-1)^l 2^{2l}}{2^{2l+1}-1}
\left\{ 
\sum_{k=1}^{l-1} \frac{(-1)^{k-1}\pi^{2(l-k)}}{(2(l-k))!}
\left( 1-\frac{1}{2^{2k}} \right) \zeta(2k+1) \right. \nonumber \\
&& \hspace{20mm} \left. 
-\frac{\pi^{2l}}{(2l)!}
\left( 
\log \pi -\frac{1}{2l}
-l\sum_{k=1}^{\infty}
\frac{\zeta(2k)}{k(k+l)2^{2k}}  \right) \right\}.
\end{eqnarray}

\par \noindent
For examples,
\begin{eqnarray*}
\zeta(3)&=&
\frac{2{\pi}^2}{7}
\left\{ 
\log \pi-\frac{1}{2}-\sum_{k=1}^{\infty}\frac{\zeta(2k)}{k(k+1)2^{2k}}
\right\}, \\
\zeta(5)&=&
\frac{6{\pi}^2}{31}
\left\{ 
\zeta(3)
-\frac{{\pi}^2}{9}
\left( \log \pi-\frac{1}{4}
-2\sum_{k=1}^{\infty}\frac{\zeta(2k)}{k(k+2)2^{2k}} \right) 
\right\} \\
&=&
\frac{4{\pi}^4}{651}
\left\{ 
\frac{11}{2}\log \pi-\frac{29}{8}-
\sum_{k=1}^{\infty}\frac{(2k+11)\zeta(2k)}{k(k+1)(k+2)2^{2k}}
\right\}. 
\end{eqnarray*}

\vspace{5mm}
A comment is in order. \ \ There are many expressions like 
(\ref{eq:main}). For example, in \cite{CK} (Theorem A) it is given  
\[
\zeta(2l+1)=(-1)^l \frac{(2\pi)^{2l}}{l(2^{2l+1}-1)}
\left[ 
\sum_{k=1}^{l-1}(-1)^{k-1} \frac{k \zeta(2k+1)}{\pi^{2k}(2l-2k)!}
+\sum_{k=1}^{\infty}\frac{\zeta(2k)(2k)!}{2^{2k}(2k+2l)!}
\right].
\]
However, our expression is different from this.

\vspace{5mm}
In this note we gave a simple method to obtain some deep relations among 
zeta--values by calculating the Euler integral (in our terminology). 
The method is systematic and most elementary as far as we know. 
Moreover, it must be fresh for not only non--experts or students but also 
experts in Elementary Number Theory. 

We conclude this note by stating our dream. We are working in some field of 
Quantum Computation, so we dream that the {\bf Riemann conjecture} will be 
finished by it.


\end{document}